\documentclass[aps,pre,amsmath,lengthcheck,superscriptaddress]{revtex4-2}

\usepackage{graphicx}\graphicspath{ {figures/} }
\usepackage{hyperref}
\hypersetup{colorlinks,allcolors=blue,breaklinks}
\usepackage{amssymb}

\newcommand{\be}{\begin{equation}}
\newcommand{\ee}{\end{equation}}
\newcommand{\ba}{\begin{align}}
\newcommand{\ea}{\end{align}}
\newcommand{\bi}{\begin{itemize}}
\newcommand{\ei}{\end{itemize}}

\newcommand{\pd}{\partial}

\newcommand{\bla}{bla\\bla\\bla\\bla\\bla}

\newcommand{\mb}[1]{\mbox{\boldmath$#1$}}
\newcommand{\mc}[1]{\mathcal{#1}}

\begin{document}

\title{Work probability distribution of weakly driven process in overdamped dynamics}

\author{Pierre Naz\'e}
\email{pierre.naze@icen.ufpa.br}

\affiliation{\it Universidade Federal do Par\'a, Faculdade de F\'isica, ICEN,
Av.~Augusto Corr\^ea, 1, Guam\'a, 66075-110, Bel\'em, Par\'a, Brazil}

\date{\today}

\begin{abstract}

Analytical work probability distributions for open classical systems are scarce; they can only be calculated in a few examples. In this work, I present a new method to derive such quantities for weakly driven processes in the overdamped regime for any switching time. The white noise Brownian motion in a harmonic linear stiffening trap illustrates the result. The work probability distribution is non-tabulated, with positive, semi-finite support, diverging at the minimal value, and non-Gaussian. An analysis of the range of validity of linear response is made by using the self-consistent criterion of the fluctuation-dissipation relation. The first, second, third, and fourth moments are correctly calculated for small perturbations.

\end{abstract}

\maketitle

\section{Introduction}

The statistical description of nonequilibrium processes has garnered significant attention over the past decades, particularly in systems driven away from equilibrium by external forces~\cite{Jarzynski1997,Crooks1999,Seifert2008}. Among the quantities of interest in such scenarios, the work performed on a system during a time-dependent process plays a central role~\cite{Speck2004,Schuler2005}
. Although equilibrium statistical mechanics provides a comprehensive framework to describe static thermodynamic quantities, obtaining analytical results for work distributions in nonequilibrium settings remains a challenging endeavor~\cite{Jarzynski2006,Esposito2009}
.

In classical systems, exact work probability distributions are known only for a few specific cases, often requiring severe simplifications such as harmonic potentials and idealized driving protocols~\cite{imparato2005work,engel2009asymptotics,kwon2013work}. The situation becomes even more restrictive for open systems in the presence of thermal noise and dissipation~\cite{Seifert2012}. In this context, linear response theory emerges as a powerful approximation tool, especially suitable for weakly driven processes~\cite{Agarwal1972,Kubo1966}
. By expanding observables to first or second order in a small perturbation parameter, linear response theory offers analytical insights into dynamical behavior without requiring full solutions to stochastic differential equations~\cite{Speck2009}.

In the case of overdamped Brownian particles in harmonic traps, linear response theory allows for explicit analytical predictions of the average work and its variance under small driving amplitudes \cite{naze2020compatibility,naze2022optimal,naze2023optimal}. These results serve as benchmarks for stochastic thermodynamics, especially when testing the limits of fluctuation theorems and nonequilibrium relations \cite{Seifert2005}. However, obtaining the full probability distribution of work analytically, even within linear response theory, remains a nontrivial task and has not receive any attention in the literature.

The focus of this work is to derive the work probability distribution for classical systems weakly driven in the overdamped regime. More specifically, I propose a method to obtain the probability distribution analytically for arbitrary switching times, provided the driving remains within the linear response domain. The formulation leverages the relation between the work and the system initial conditions, transforming the problem into one of variable change in the probability measure.

To illustrate the generality and effectiveness of the method, I consider the paradigmatic example of an white noise overdamped Brownian particle, subjected to a harmonic linear stiffness trap. This example is particularly appealing, as it combines analytical tractability with physical relevance~\cite{bustamante2005nonequilibrium,schmiedl2007optimal,Seifert2012}. Comparisons with numerically exact simulations confirm the accuracy of the predictions, including a detailed analysis of the first four moments of the distribution. Furthermore, the behavior of the work distribution under varying switching times and driving amplitudes reveals how the system transitions from linear to nonlinear response. In particular, I present a self-consistent method based on the fluctuation-dissipation relation~\cite{naze2023optimal}
\be
\langle \overline{W}_\tau\rangle = \Delta F +\frac{\beta}{2}\sigma_{W_\tau}^2,
\label{eq:fdr}
\ee
where no extra calculation of next order is necessary to determine the range of validity of linear response theory. Here, $\langle \overline{W}_\tau\rangle$ is the averaged work, $\Delta F$ the difference in Helmholtz's free energies and $\sigma_{W_\tau}^2$ the variance of work.

This manuscript is organized as follows. First, I introduce the theoretical framework and derive the general expression for the work distribution. Next, I apply the method to the stiffening trap model and obtain the probability distribution function. Then, I perform a thorough comparison between the linear response prediction and exact numerical results, including an evaluation of the range of validity of the method. Finally, I conclude with a discussion of the implications of the results and potential directions for future extensions.

\section{Weakly driven processes}
\label{sec:lrt}

I start by defining notations and developing the main concepts to be used in this work. 

Consider a classical system with a Hamiltonian $\mc{H}(\mb{z}(\mb{z_0},t)),\lambda(t))$, where $\mb{z}(\mb{z_0},t)$ is a point in the phase space $\Gamma$ evolved from the initial point $\mb{z_0}$ until time $t$, with $\lambda(t)$ being a time-dependent external parameter. Initially, the system is at equilibrium with a heat bath of temperature $\beta\equiv {(k_B T)}^{-1}$, where $k_B$ is Boltzmann's constant. During a switching time $\tau$, the external parameter is changed from $\lambda_0$ to $\lambda_0+\delta\lambda$, with the system always in contact with the initial heat bath. The configuration of the system is such that the driving is made in the overdamped regime, where the acceleration is negligible. The work performed on the system during this interval of time is
\be
\overline{W}_\tau(x_0) \equiv \int_0^\tau \overline{\pd_{\lambda}\mc{H}}(x_0,t)\dot{\lambda}(t)dt,
\label{eq:work}
\ee
where $\partial_\lambda$ is the partial derivative in respect to $\lambda$ and the superscripted dot the total time derivative. Since the work is a random variable, the generalized force $\overline{\pd_{\lambda}\mc{H}}(x_0,t)$ depends on the initial position $x_0$, which obeys the canonical ensemble $\rho_{x_0}(x_0)$. The overline symbol $\overline{\cdot}$ denotes an average over the noise produced by the heat bath. The external parameter can be expressed as
\be
\lambda(t) = \lambda_0+g(t)\delta\lambda,
\label{eq:ExternalParameter}
\ee
where to satisfy the initial conditions of the external parameter the protocol $g(t)$ must satisfy the following boundary conditions $g(0)=0$, $g(\tau)=1$.

Linear-response theory aims to express average quantities until the first-order of some perturbation parameter considering how this perturbation affects the observable to be averaged and the probabilistic distribution \cite{kubo2012}. In our case, I consider that the parameter does not considerably change during the process, $|g(t)\delta\lambda/\lambda_0|\ll 1$, for all $t\in[0,\tau]$ and $\lambda_0\neq 0$. Also, the average of the initial conditions does not need to be performed. The work can be approximated up to second-order as
\begin{equation}
\begin{split}
\overline{W}_\tau(x_0) = &\, \delta\lambda\overline{\pd_{\lambda}\mc{H}}(x_0)-\frac{\delta\lambda^2}{2}\widetilde{\psi}(x_0)\\
&+\frac{1}{2}\int_0^\tau\int_0^\tau \psi(x_0,t-t')\dot{\lambda}(t')\dot{\lambda}(t)dt'dt,
\label{eq:work2}
\end{split}
\end{equation}
where 
\be
\psi(x_0,t) = \beta\pd_\lambda\mc{H}(x_0,0)\overline{\pd_\lambda\mc{H}}(x_0,t)-c(x_0)
\ee 
is the (not-averaged) relaxation function, with $\widetilde{\psi}_0(x_0)\equiv \psi_0(x_0,0)-\overline{\pd_{\lambda\lambda}^2\mc{H}}(x_0,0)$ \cite{kubo2012}. The constant $c(x_0)$ is chosen to nullify the (averaged) relaxation function for long times. 

Such an equation holds for finite-time and weakly driven processes. In particular, the averaged work could be split into the difference of Helmholtz's free energy and irreversible work, given by
\be
\Delta F(x_0) = \delta\lambda\overline{\pd_{\lambda}\mc{H}}(x_0)-\frac{\delta\lambda^2}{2}\widetilde{\psi}(x_0),
\label{eq:dF}
\ee
\be
W^{\rm irr}_\tau(x_0) = \frac{1}{2}\int_0^\tau\int_0^\tau \psi(x_0,t-t')\dot{\lambda}(t')\dot{\lambda}(t)dt'dt.
\label{eq:wirr}
\ee
Observe that both quantities depend on the initial position $x_0$. Our objective is to derive the work probability distribution and to corroborate the analytical prediction with a comparison with the exact result for the white noise overdamped Brownian motion, subject to a harmonic linear stiffening trap.

\section{Work probability distribution}

To find the work probability distribution, one needs to make a change of variables in the initial position probability distribution, considering the inverted function
\be
x_0 = x_0(\overline{W}_\tau,\tau).
\ee
In this manner
\be
\rho_{x_0}(x_0)dx_0 = \rho_{x_0}(x_0(\overline{W}_\tau,\tau))\left|\frac{dx_0}{d\overline{W}_\tau}\right|d\overline{W}_\tau.
\ee
Therefore, the work probability distribution is
\be
\rho_{\overline{W}_\tau}(\overline{W}_\tau) = \rho_{x_0}(x_0(\overline{W}_\tau,\tau))\left|\frac{dx_0}{d\overline{W}_\tau}\right|.
\ee
Such expression holds in principle for all switching times $\tau$ and, in case the inverse is not unique, a sum must be made with all solutions. The difficulty lies in the inversion of the function $\overline{W}_\tau(x_0)$, which is not always easy to analytically obtain.

\section{Example: stiffening trap}

I consider a white noise overdamped Brownian motion subjected to a time-dependent harmonic potential, with the mass of the system equal to one, $\gamma$ as a damping coefficient, and $\omega_0$ as the natural frequency of the potential. The not-averaged relaxation function is given by
\be
\psi(x_0,t)=\frac{\beta}{4}\left(x_0^4-\frac{x_0^2}{\beta\lambda_0}\right)\exp{\left(-\frac{2\omega_0^2}{\gamma}|t|\right)}.
\ee
The Brownian particle is subject to a harmonic linear stiffening trap, given accordingly to the following protocol
\be
g(t)=\frac{t}{\tau}.
\ee
Such an example is relevant due to its analytical tractability and for being used as a testbed in experiments.

\subsection{Probability distribution function}

The work calculated by linear response theory is invertible in $x_0$ with four solutions, with two in real numbers. In this manner, the expression for the work probability function has the following aspect
\be
\rho_{\overline{W}_\tau}(w) \propto 
\left|\frac{
\exp\left[ -\frac{1}{2} \left( A - B \sqrt{a - b w} \right) \right]
}{
\sqrt{A - B \sqrt{a - b w}} \cdot \sqrt{a - b w}
}\right|,
\ee
where the coefficients depend on the parameters $\delta\lambda/\lambda_0$ and $\tau/\tau_R$. This expression is not recognizable as a tabulated probability distribution function. It also shows that the support is positive, semi-finite and diverges with $1/\sqrt{a-b w}$ at the minimal work value $a/b$. Indeed, for determined $\tau/\tau_R$ and $\delta\lambda/\lambda_0$, the most visited work is the minimal one. This also says that the averaged work has an optimal value~\cite{naze2024analytical}. Last but not least, the behavior is not Gaussian since it exponentially decays with the square root of the work.

\subsection{Comparison with exact histogram}

\begin{figure}[t]
    \centering
    \includegraphics[width=0.75\linewidth]{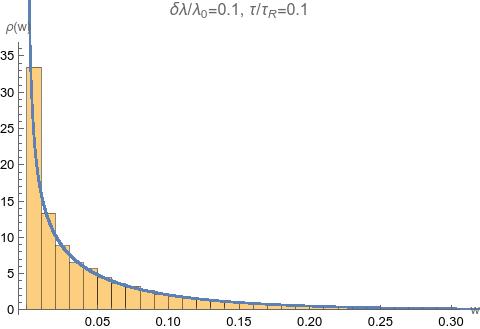}
        \caption{Work probability distribution for white noise overdamped Brownian motion under harmonic linear stiffening trap. It was used $\tau/\tau_R=0.1$, for $\delta\lambda/\lambda_0=0.1$. Histogram and linear response prediction match.}
    \label{fig:1}
\end{figure}

\begin{figure}[t]
    \centering
    \includegraphics[width=0.75\linewidth]{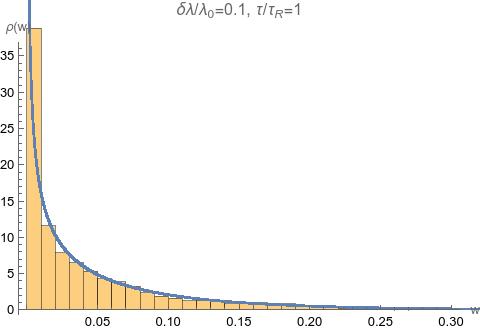}
    \caption{Work probability distribution for white noise overdamped Brownian motion under harmonic linear stiffening trap. It was used $\tau/\tau_R=1$, for $\delta\lambda/\lambda_0=0.1$. Histogram and linear response prediction match.}
    \label{fig:2}
\end{figure}

\begin{figure}[t]
    \centering
    \includegraphics[width=0.75\linewidth]{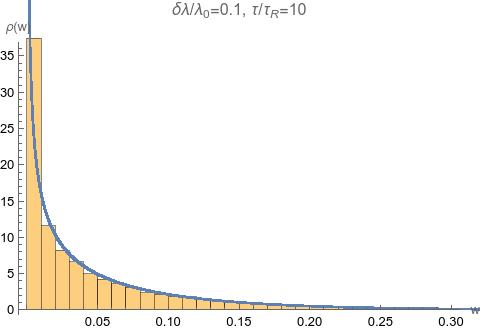}
    \caption{Work probability distribution for white noise overdamped Brownian motion under harmonic linear stiffening trap. It was used $\tau/\tau_R=10$, for $\delta\lambda/\lambda_0=0.1$. Histogram and linear response prediction match.}
    \label{fig:3}
\end{figure}

To evaluate the validity of the result, the exact work has been calculated with software to be used as a benchmark in comparisons with linear response theory. First, all simulations were tested to see if they satisfy Jarzynski's equality~\cite{Jarzynski1997}, being successful in all attempts. It was performed $N=10^4$ processes for each one of the switching times $\tau/\tau_R=0.1,1,10$ with $\delta\lambda/\lambda_0=0.1$. In Fig.~\ref{fig:1},~\ref{fig:2},~\ref{fig:3}, I present the match between the numerical solution of the simulations and the prediction by linear response theory. It is curious that the distribution almost does not change varying the switching time; this indicates that the weak perturbation reduces the effect of such a change. Indeed, the parameter $\tau/\tau_R$ only appears in the irreversible work (see Eq.~\eqref{eq:wirr}), which is of second-order in $\delta\lambda/\lambda_0$. The asymmetry of the graphics indicates the non-Gaussian aspect of the distribution.

\subsection{Range of validity analysis}

\begin{figure}[t]
    \centering
    \includegraphics[width=0.75\linewidth]{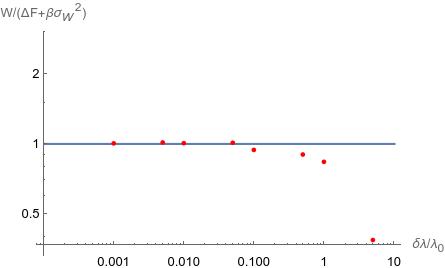}
    \caption{Verification of the fluctuation-dissipation relation for different $\delta\lambda/\lambda_0$. For strong perturbations ($\gtrsim 0.1$), linear response breaks down.}
    \label{fig:5}
\end{figure}

In Fig.~\ref{fig:5}, to analyze the range of validity of linear response theory, I present the case with parameters $\tau/\tau_R=1$ and $\delta\lambda/\lambda_0=0.5$. The strong intensity clearly breaks down linear response theory. To see a more quantitative method to evaluate the correspondence of the exact result with the linear response approximation is by verifying the fluctuation-dissipation relation (see Eq.~\eqref{eq:fdr}). Indeed, such relation must hold only in linear response theory~\cite{naze2023optimal}. For $\tau/\tau_R=1$, I observe that perturbations greater than $\delta\lambda/\lambda_0=0.1$ linear response approximation breaks down. This indicates that the strong driving puts the system in a far-from-equilibrium regime.

\begin{figure}[t]
    \centering
    \includegraphics[width=0.75\linewidth]{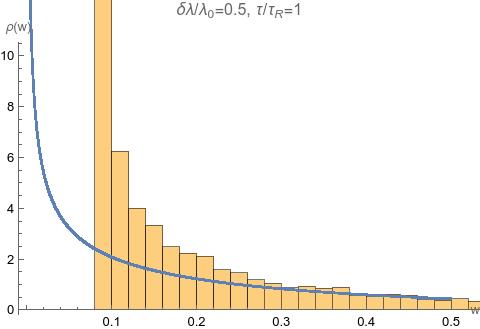}
    \caption{Work probability distribution for white noise overdamped Brownian motion under harmonic linear stiffening trap. It was used $\tau/\tau_R=1$, for $\delta\lambda/\lambda_0=0.5$. Linear response theory breaks down. This indicates that the strong driving puts the system in a far-from-equilibrium regime.}
    \label{fig:4}
\end{figure}

\subsection{Self-consistent criterion}

One of the main problems in verifying the range of validity of linear response theory is the necessity of having the exact solution at hand, or, at least, the next order calculated. I present now a self-consistent criterion of evaluation of the range of validity of linear response based only on the work probability distribution calculated by this approximation theory. 

\begin{figure}[t]
    \centering
    \includegraphics[width=0.75\linewidth]{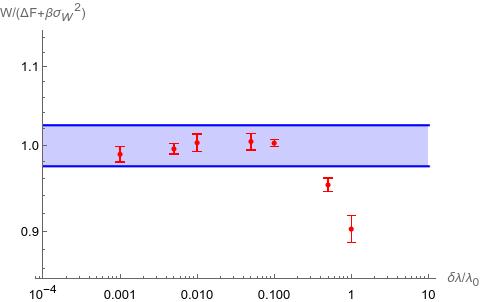}
    \caption{Verification of the fluctuation-dissipation relation for different $\delta\lambda/\lambda_0$. For strong perturbations ($\gtrsim 0.1$), linear response breaks down. This is a self-consistent check for the range of validity of linear response theory.}
    \label{fig:6}
\end{figure}

Using again the fluctuation-dissipation relation in the average and variance calculated by samples of the work probability distribution, I observe that the relation should hold only for small perturbations. Indeed, with the whole distribution, higher order cumulants, like the skewness and kurtosis, participate in the relation between the average and the variance, becoming the fluctuation-dissipation relation no longer holding. Observe that this is only possible using the work probability distribution since the direct calculation of the average and variance will always satisfy the desired relation. Figure~\ref{fig:6} corroborates our intuition. The simulations were done in the same conditions of Fig.~\ref{fig:5}, illustrating that for the same strong perturbations observed before ($>0.1$), linear response theory breaks down. This self-consistent criterion is of utmost importance for linear response theory since no extra calculations will be necessary to verify its range of validity.

\subsection{Moment analysis}

\begin{figure}[t]
    \centering
    \includegraphics[width=0.75\linewidth]{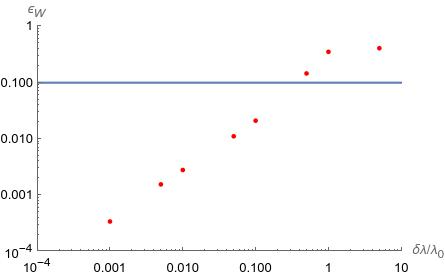}
    \caption{Comparison of the average work of the exact work and the prediction of linear response theory. For strong perturbations ($\gtrsim 0.1$), linear response breaks down. It was used $\tau/\tau_R=1$.}
    \label{fig:7}
\end{figure}

\begin{figure}[t]
    \centering
    \includegraphics[width=0.75\linewidth]{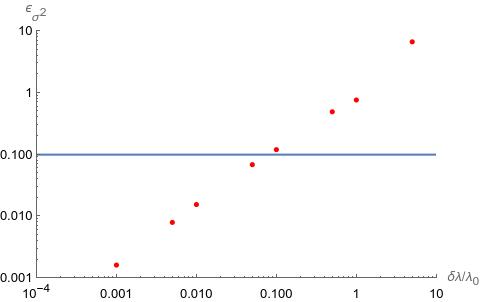}
    \caption{Comparison of the work variance of the exact work and the prediction of linear response theory. For strong perturbations ($\gtrsim 0.1$), linear response breaks down. It was used $\tau/\tau_R=1$.}
    \label{fig:8}
\end{figure}

\begin{figure}[t]
    \centering
    \includegraphics[width=0.75\linewidth]{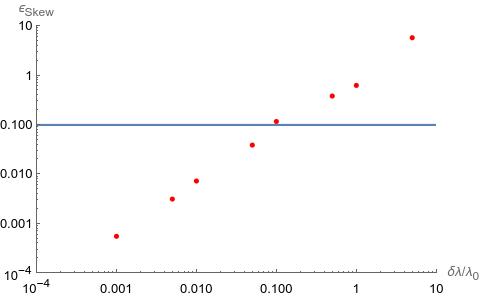}
    \caption{Comparison of the work skewness of the exact work and the prediction of linear response theory. For strong perturbations ($\gtrsim 0.1$), linear response breaks down. It was used $\tau/\tau_R=1$.}
    \label{fig:9}
\end{figure}

\begin{figure}
    \centering
    \includegraphics[width=0.75\linewidth]{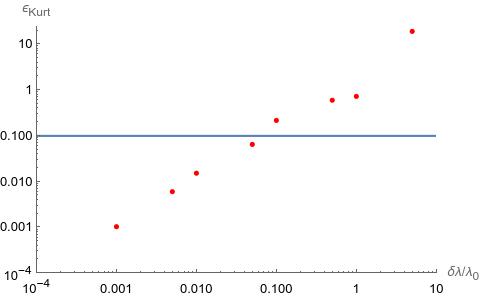}
    \caption{Comparison of the work kurtosis of the exact work and the prediction of linear response theory. For strong perturbations ($\gtrsim 0.1$), linear response breaks down. It was used $\tau/\tau_R=1$.}
    \label{fig:10}
\end{figure}

To evaluate if the work probability distribution furnishes the first, second, third, and fourth moments, I compare the relative error of the average and variance of the exact distribution and the one predicted by linear response theory. Figures~\ref{fig:7},~\ref{fig:8},~\ref{fig:9}, and~\ref{fig:10} respectively illustrate the relative errors for the average, variance, skewness, and kurtosis for $\tau/\tau_R=1$ and different perturbations. The prediction has relative errors of less than $10\%$ for perturbations less than $\delta\lambda/\lambda_0=0.1$. Indeed, the relation shows a linear relative error increase with the perturbation. This indicates that the strong driving puts the system in a far-from-equilibrium regime. Also, the existence of skewness and kurtosis more quantitatively indicates that the probability distribution is not Gaussian.

\section{Final remarks}
\label{sec:final}

In this work, I presented an analytical method to obtain the work probability distribution for open classical systems performing weak drivings in the overdamped regime. The method is exemplified by the white noise overdamped Brownian motion, which is subjected to a harmonic linear stiffening trap for different switching times. The work probability distribution is non-tabulated, with positive, semi-finite support, diverging at the minimal value, and non-Gaussian. An analysis of the range of validity of linear response is made by using the self-consistent criterion of the fluctuation-dissipation relation. The first, second, third, and fourth moments are correctly calculated for small perturbations. 

Natural extensions of the method are for the underdamped regime and thermally isolated systems, where the change of variable will occur for two variables now. Understanding the role of optimality concerning the optimal protocols~\cite{naze2024analytical} in the work probability distribution is an important point to explore as well. Reaching a better response to strong driving requires extension to nonlinear response theory, which is still a demand. All these topics will be studied in future research. 

\section*{Data availability} 

The code used in the simulations can be found at \hyperlink{https://github.com/pnaze/PDLR/}{https://github.com/pnaze/PDLR/}.

\raggedbottom

\bibliography{PDLR.bib}
\bibliographystyle{apsrev4-2}

\end{document}